\documentclass[useAMS,usenatbib]{mn2e}

\usepackage{graphicx}

\newcommand{\e}[1]{\ensuremath{{}_{\rm{#1}}}}
\newcommand{\U}[1]{\ensuremath{\mathrm{~#1}}}
\newcommand{\Myr}{\U{Myr}}
\newcommand{\kpc}{\U{kpc}}
\newcommand{\solarm}{\U{M}_{\odot}}

\title{Star cluster survival and compressive tides in Antennae-like mergers}

\author[Renaud et al.]{F.~Renaud$^{1,2}$ C.~M.~Boily$^{1}$, J.-J.~Fleck$^{1}$, T.~Naab$^{3}$ and Ch.~Theis$^{2}$\\
$^{1}$Observatoire astronomique and CNRS UMR 7550, Universit\'e de Strasbourg, 11 rue de l'Universit\'e, F-67000 Strasbourg, France \\
$^{2}$Institut f\"ur Astronomie der Univ. Wien, T\"urkenschanzstr. 17, A-1180 Vienna, Austria \\
$^{3}$University Observatory, Scheinerstr. 1, D-81679 Munich, Germany}

\date{Accepted 15 September 2008}

\begin{document}

\maketitle

\begin{abstract}
Gravitational tides are widely understood to strip and destroy galactic substructures. In the course of a galaxy merger, however, transient totally compressive tides may develop and prevent star forming regions from dissolving, after they condensed to form clusters of stars. We study the statistics of such compressive modes in an N-body model of the galaxy merger NGC~4038/39 (the Antennae) and show that $\simeq 15\%$ of the disc material undergoes compressive tides at pericentre. The spatial distribution of observed young clusters in the overlap and nuclear regions of the Antennae matches surprisingly well the location of compressive tides obtained from simulation data. Furthermore, the statistics of time intervals spent by individual particles embedded in a compressive tide yields a log-normal distribution of characteristic time $\tau \sim 10 \Myr$, comparable to star cluster formation timescales. We argue that this generic process is operative in galaxy mergers at all redshifts and possibly enhances the formation of star clusters. We show with a model calculation that this process will prevent the dissolution of a star cluster during the formation phase, even for a star formation efficiency $\epsilon$ as low as $\sim 10\%$. The transient nature of compressive tides implies that clusters may dissolve rapidly once the tidal field switches to the usual disruptive mode.
\end{abstract}


\section{Introduction}

Galaxy interactions are known to trigger star cluster formation. A typical example is NGC~4038/39 (the Antennae), where some 1000 star clusters has been detected \citep{Mengel05,Whitmore07}. However, it is unclear how many of them will survive internal destruction: for instance, low star formation efficiency (SFE) or strong stellar feedback could destroy the molecular cloud \citep{Elmegreen97}) and lead to ``infant-mortality'' \citep{Bastian06,Whitmore07,deGrijs07}. Furthermore, these questions are central to interpreting the enhanced specific frequency and metallicity bi-modality of clusters in giant ellipticals \citep{Harris91}. In this Letter, we stress that totally \emph{compressive} tides widely exist in galaxy mergers and slowdown the early dissolution of clusters.

HST V-band and deep IR images of the Antennae galaxies have suggested a near power-law cluster mass function ($\psi[M] \propto M^{-2}$ for $10^4 \solarm < M < 10^6 \solarm$; \citealt{Fall05,Anders07}) and an age distribution with a rough power-law fit ($dN/d\tau \propto \tau^{-1}$) suggesting that many young clusters dissolve rapidly after they are born inside their primordial host gas cloud \citep{Fall05}.

The evolution of (broad-band) mass-to-light ratios offers a strong hint that young clusters are severely perturbed and may even be out of dynamical equilibrium \citep{Bastian06,deGrijs07}. Yet the low inter-cluster velocity dispersion measured in the Antennae galaxy merger indicates that cluster formation proceeds from a coherent coalescence of large-scale material as opposed to \emph{in situ} shock dissipation and fragmentation \citep{Whitmore05}. Computer simulations of galaxy mergers have long suggested that streams of material (made up of both gas and stars) developing during a merger will form dense structures, so increasing the star- and cluster-formation rates \citep{Barnes88,Barnes92,Mihos96,Naab06}. A Schmidt law or variation thereof can be applied to circumvent numerical resolution issues and derive estimates of the star formation rate in rough agreement with observations (see e.g. \citealt{Barnes04} for an application to the Mice; \citealt{diMatteo07}).

Despite this wealth of progress, the long-term fate of star forming regions (bound structures or not) remains largely unclear, especially with regard to star cluster complexes such as those of the Antennae galaxies. The situation is exacerbated in the case of a galaxy merger because the gravitational tides vary rapidly in time along with the morphology of the system as a whole. An observational snapshot showing a candidate star-forming aggregate lacks information about the duration and motion in space to determine whether the aggregate will remain bound or not. When matter (gas or stars) is distributed in space in fragmented and irregular fashion, as is typical of galaxy mergers, the tidal field may exert either a disruptive outward stretch on an extended mass; or, on the contrary, it may exert an inward compressive force, which then cocoons the volume enclosed and increases its binding energy. While converging streams have long been suspected of driving the formation of tidal dwarf galaxies \citep[e.g.][]{Barnes92,Bournaud06,Wetzstein07}, the impact of compressive tides themselves on sub-galactic scales has never been considered strictly as a gravitational effect. Previous authors analysed cluster formation as the result of a global trigger, using sticky-particles \citep{Bournaud08} or SPH simulations with a threshold in pressure \citep{Bekki02a,Bekki02b} or in density \citep{Li04} for the formation of stars.

Here, we take a different stand by exploring the impact of compressive tides on cluster-size volume elements. We present in \S2 a high-resolution numerical model of the Antennae galaxy merger to match kinematic and optical data. In \S3, we discuss the key results obtained from the computer simulation. We close with a discussion of how the process outlined here bears on the statistics of observed cluster populations.

\section{Merger simulation}
We picked two equal-mass bulge/disc/halo galaxy models to simulate an Antennae-like major merger similar to that of \citet{Barnes88}. We set up self-gravitating models using the {\tt magalie} software included with the Nemo\footnote{http://bima.astro.umd.edu/nemo/} stellar dynamical package \citep[version 3.2.4]{Teuben95}. The equations of motion were integrated with gyrfalcON \citep{Dehnen02} using a Plummer softening kernel with $b = 0.01$ in units where the gravitational constant $G = 1$, a unit of time $= 25 \Myr$, a mass unit $= 3.6 \times 10^{10} \solarm$ and a length unit of $4.4 \kpc$. The total mass of each galaxy is equal to 7 mass units, where the exponential disc scale-length and mass $r_d = M_d = 1$. The bulge is a \citet{Hernquist90} model with a scale parameter $a = 1$ and total mass\footnote{Our model galaxies differ mainly from Barnes's in that he used a lighter \citet{King66} model for the bulge.} $M_b = 1$. Thus our progenitor galaxies are ``S0 - Sa'' galaxies with a Toomre parameter $Q = 1.5$ which ensures local stability to axially symmetric fragmentation modes. An isothermal dark matter halo of total mass $M_h = 5$, of equal truncation radius and scale length $r_{h,t} = r_h = 7$, was set up as in \citet{Hernquist93}. The mass ratio between individual component particles of $\sim 2$ keeps two-body heating low. A total of $700,000$ particles were used for each galaxy, for a mass resolution of $3.50$, $1.75$ and $4.38 \times 10^5 \solarm$ respectively for each component\footnote{As the mass of the Antennae is measured with less accuracy than the velocities, we derived a mass-scale from $\sigma^2 = GM/r$ after adjusting the kinematics.}. Lower resolution simulations with 175,000 and 350,000 bodies per galaxies did not change significantly the statistics reported here, demonstrating that the results are robust to Poisson noise. No gas was included in our simulation. The typically small gas mass fraction would not affect the gravitationnal potential significantly. Hydrodynamical aspects would be more important at the star-formation-scale, where e.g. fragmentation, winds and supernovae explosions play a crucial role.

The galaxies were set on a prograde encounter with tidal tails developing on either side of the orbital plane, implying non-coplanar discs\footnote{See movies at http://astro.u-strasbg.fr/$\sim$renaud/simu.php}. The orbit of the associated point-mass problem is bound with eccentricity $e \approx 0.96$. The extended massive dark matter halos cause the galaxies to brake significantly already at  the first pericentre passage ($t \approx 11$, snapshot used as our time reference). Tidal tails form soon after the first passage and continue to expand until the end of the simulation at $t \approx 30$. Comparing with HI maps of \citet{Hibbard01} and optical images of the Antennae, the merger simulation gives a good match to the morphology and kinematics (peak systemic velocity of 130 km/s) at time $t \approx 23$, when a second passage through pericentre is in full swing (see Renaud et al. 2009, in preparation, for details).

\section{Statistical approach and results}
Because the gas from which clusters form is largely confined to the disc of spirals, we focus on the fate of disc particles. The mass resolution of $\sim 3 \times 10^5 \solarm$ allows to treat individual disc particles as cluster-size elements\footnote{Resolution requirements do not permit to simulate an individual star clusters on the scale of a galaxy merger.}.

This resolution allows to monitor precisely the tidal field around a cluster. To study the effect of tides on a hypothetical star inside a cluster, let the cluster be at the origin of an inertial barycentric reference frame. The net force of the star is the sum of internal cluster and external galactic gravity. The tidal tensor computed from this external acceleration reads $T^{ij}=\partial g\e{ext}^i/\partial x^j$, where $x^j$ is the $j$-th component of the position vector and $g\e{ext}$ is the acceleration. We retrieved the six-component tidal tensor of the $4 \times 10^5$ discs particles at each snapshot output time with a second-order finite difference scheme applied to a cube of side $b = 40 \U{pc}$. The symmetric tensor $T$ admits a diagonal form with eigenvalues $\lambda_1 \leq \lambda_2 \leq \lambda_3$ computed using the method of \citet{Kopp06}. The signs of the eigenvalues determine if the differential force leads to a stretch ($\lambda > 0$) or a compression ($\lambda < 0$) of the model cluster. Most examples of analytic potentials with a cored density show a mixture of both cases \citep{Valluri93,Dekel03,Gieles07}. From here onwards, a fully compressive tidal mode is understood to be one where all three eigenvalues are negative.
\begin{figure}
\includegraphics[width=84mm]{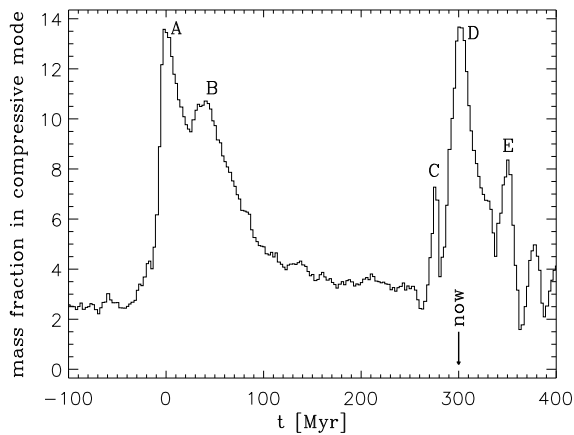}
\caption{Percentage of the disc mass in compressive mode as a function of time. The peaks labelled 'A' through 'E' correspond to pericentre passages (A,D) and rapid morphological evolution (B,C,E), respectively. See text for details.}
\label{fig:histo}
\end{figure}
\begin{figure*}
\includegraphics[width=168mm]{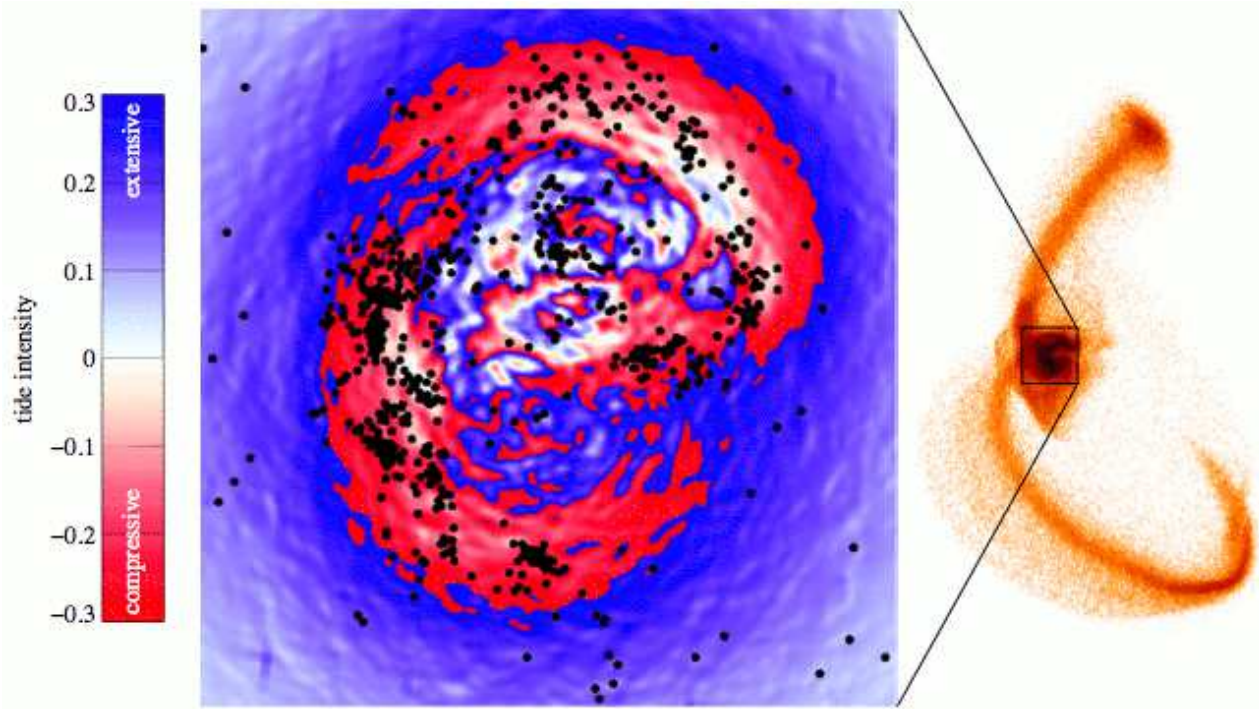}
\caption{Left: map of gravitational tides in the nuclear region of the simulated merger ($\sim 12 \kpc$ wide). The black dots are candidate clusters identified by \citet{Mengel05} from VLT images. The blue to red colours label extensive to progressively more compressive tidal modes. The two nuclei are marked by green crosses. Right: large-scale column-density map of the simulation at $t\simeq 23$ time units ($\sim 300 \Myr$, peak 'D' in Fig.~\ref{fig:histo}).}
\label{fig:map}
\end{figure*}
In Fig.~\ref{fig:histo}, we plot the percentage of disc particles in compressive mode as a function of time. The main evolutionary stages of the merger can easily be identified from the rapid increase in particles undergoing compression. Peak 'A' occurs at the first pericentre passage, when the progenitors are largely whole, and the tidal field at its strongest. Around that time, bridges of matter form between the two progenitors and cause a second peak 'B' of compressive modes, of a somewhat longer duration. The number of compressive regions decreases steadily as the galaxies move apart, in the interval $50 < t < 250 \Myr$, prior to the second passage to pericentre.  Once again a strong wave of compressive modes brings $\sim 15\%$ of the bodies to experience compressive tides (peak 'D' on Fig.~\ref{fig:histo}). Numerous short-duration spikes are visible around this second encounter (denoted 'C' and 'E' on the figure) which are linked to streams of tidal clumps falling onto the newly formed core.

Roughly speaking, the mass fraction in tidal compression rises five fold when compared to the progenitor galaxies in isolation. This relative increase compares well with  hydrodynamical simulations of a major merger, where the number of star forming regions also increases on average five times at pericentre. \Citet{diMatteo08} report that $80 \%$ of a sample of $\sim 900$ GalMer merger simulations showed a relative star formation rate increase of $\leq 5$. Larger increases would require a specific configuration, or more effective gas dissipation \citep[see e.g.][]{Mihos96,Springel00}.

Fig.~\ref{fig:map} maps the spatial distribution of gravitational tidal modes of the nuclear region using a colour-coded scheme with blue meaning extensive tides, and red compressive ones. A projection of the particles on the sky is shown to the right for reference. The map was constructed with an $26.4 \kpc \times 26.4 \kpc$ grid of $256^2$ mesh points. Large spiral patterns of compressive modes are clearly visible and reminiscent of the distribution of massive young clusters dispatched throughout the tidal bridge linking the two nuclei of the Antennae \citep[the overlap region; see Fig.~1 of][]{Mengel05}. An overlay of the tidal map to a grey-scale rendition of Mengel et al.'s image shows a clear correlation between the distribution of clusters (shown as dots) and the morphology and intensity of the compressive tides. The agreement with observations is especially good for the northern arc-like feature and throughout the overlap region.
\begin{figure}
\includegraphics[width=84mm]{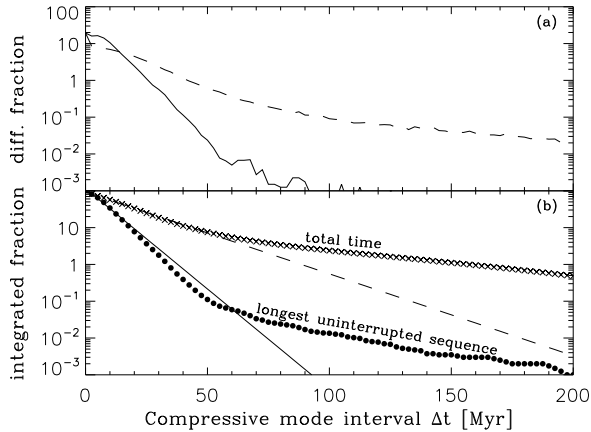}
\caption{Distributions of the time-intervals $\Delta t$ spent in compressive modes by disc particles. (a) top panel: differential fraction ${\cal M}(\Delta t+dt) - {\cal M}(\Delta t)$; (b) bottom panel: forward-integrated cumulative fraction ${\cal M}(>\Delta t)$. The solid curves show the distributions of the longest uninterrupted sequences ($lus$), while the dash is for the total time ($tt$) in compressive mode.}
\label{fig:period}
\end{figure}
To determine whether the coincidence of young clusters in regions of compressive tides is a chance event, we investigated the distribution of their duration along individual orbits. Two statistical distributions were compiled, one for the Total Time ($tt$ for short) spent in compressive tides during the whole simulation; the other covering the Longest Uninterrupted Sequence ($lus$ for short) of snapshots with a particle immersed in a compressive tide.

The results are graphed on Fig.~\ref{fig:period} in log-normal axes. The top panel on the figure shows the differential mass fraction for both distributions, ${\cal M}(\Delta t+dt)-{\cal M}(\Delta t)$, versus the time interval $\Delta t$; here $dt = 2.5 \Myr$ is the snapshot sampling rate. The bottom panel shows the forward-integrated distributions ${\cal M}(>\Delta t)$, the mass fraction of particles experiencing a compressive tide of duration exceeding $\Delta t$, for each distribution. The four curves on Fig.~\ref{fig:period} show a knee at $\Delta t \approx 50 \Myr$ ($lus$ distributions) and $80 \Myr$ ($tt$ distributions). Up to these times, the distributions are well-fitted with exponential decays ${\cal M}(\Delta t) \propto \exp{(-\Delta t/\tau)}$ of characteristic times $\tau \approx 8 \Myr$ ($lus$ distribution) and $\tau \approx 20 \Myr$ ($tt$ distributions; cf. straight lines on Fig.~\ref{fig:period}b). The interval $\tau \approx 10 \Myr$ far exceeds the formation time of O/B stars; what is more, about $40 \%$ of the $lus$ sample experiences a compressive mode of duration longer than $\tau = 10 \Myr$. The characteristic decay time $\tau \approx 20 \Myr$ of the $tt$ sample comfortably covers the formation phase of rich star clusters, including the residual-gas evacuation process.

Because the model simulation discussed here involves gravitational dynamics only, a re-scaling leaving the dimensionless quantity $GM/rv^2$ unchanged will give exactly the same results. Therefore, any major merger of mass unit $M^\prime$ and length unit $r^\prime$, will yield the same characteristic time-scale $\tau \sim 10 \U{Myr}$ for the statistics of compressive tides provided the ratio $(M^\prime/M)/(r^\prime/r)^3 = $ constant (cf. \S2 for the units of $M$ and $r$). The same applies to the eigenvalues $\lambda$ of the tidal tensor. 

\section{Discussion}
The role of tidal fields in galaxy mergers is traditionally taken to be disruptive and to inhibit star formation. The analysis of a major merger simulation suggests instead that transient morphological features give rise to compressive modes of gravitational tides on scales $\sim 40 \U{pc}$, comparable e.g. to the cut-off radius of Milky Way clusters. The characteristic times $\tau \approx 10$ to $20 \Myr$ derived for the log-normal distributions of duration interval $\Delta t$, ${\cal M}(\Delta t)$,  serves to estimate the fraction of young Antennae clusters still embedded in compressive tides at the current time. When normalised so ${\cal M}(0) = 1$, the quantity $1 - {\cal M}(\Delta t = \tau) \simeq 36\%$ gives the fraction of clusters of ages $\sim 10 \Myr$ or less still confined to a region of compressive tide. These clusters would appear super-virialised which may explain the scatter seen in mass-to-light ratio among the younger population of Antennae clusters \citep{Bastian06,deGrijs07}. Note that these clusters would all have formed during the second passage through pericentre (Fig.~\ref{fig:histo}), and the vast majority of them will soon emerge from their pressurised ``cocoon'' and experience the usual disruptive effect of the tidal field, causing a rapid increase in the dissolution rate of clusters (see the sharp drop around peak D, Fig.~\ref{fig:period}). The statistics of compressive tides also imply that some two-thirds of clusters with ages $> 30 \Myr$ are now in extensive tides and may have suffered significant mass loss or total disruption: as a result the age distribution of young clusters should peak sharply around ages $\sim 10 \Myr$, causing a break in the power-law fit of \citet{Fall05}. This conclusion is not inconsistent with their data for clusters in the mass range $< 10^5 \solarm$.

To contrast quantitatively the effect of tides on a star-forming region with the case of an isolated molecular cloud, we define the dimensionless ratio $A \equiv \alpha \lambda {R_v}^2\, R_v / ( G M\e{tot} )$ of tidal to gravitational binding energy of the cluster-forming cloud; $\lambda$ is once more the eigenvalue of a tidal tensor which is taken to be isotropic for simplicity, $R_v$ the virial radius of the cloud of total mass $M\e{tot}$, and $\alpha \sim 1$ a dimensionless quantity which depends on the mass distribution of the model cloud (e.g. $\alpha = 0.86$ for a homogeneous sphere). Because a fraction $\epsilon < 1$ of the total cloud mass is turned into stars, a young cluster forms stars while still embedded in residual gas. The ejection of this gas through feedback from stellar radiation may cause it to disrupt \citep{Goodwin97,Geyer01,Kroupa01}. A classic result is that this will occur whenever $\epsilon < 1/2$ \citep{Hills80,Goodwin97,Boily03a}. Repeating the derivation of \citet{Hills80} but with a background tidal field, we find a new equilibrium radius $R \equiv r R_v$ after gas expulsion, where $r$ is obtained from
\begin{equation}
2 A \ r^3 + \left( \frac{1}{2} - \epsilon - 2 A \right) r + \frac{\epsilon}{2} = 0 \ . \label{eq:a}
\end{equation}
A solution $r > 0$ still exists when $\epsilon < 1/2$ for all $A < 0$, corresponding to compressive tides. By contrast, a solution with $A > 0$ requires $\epsilon > 1/2$ or more efficient star formation, as anticipated for a disruptive tidal mode. (The case $A = 0$ gives the critical value $\epsilon = 1/2$ for isolated clusters.) Crucial to cluster survival, a compressive tide will hold together a star-forming region that would rapidly dissolve in isolation. This remains true even for a very low star-formation efficiency $\epsilon$.

\citet{Boily03b} performed N-body calculations to address this question without any tidal field. Using the same numerical set-up and an additional compressive tide with $A = (\alpha /5) \lambda \simeq -0.06$, \citet{Fleck07} found a new equilibrium after expansion by a factor of $\simeq 2$, for an SFE $\epsilon \approx 0.10$. Furthermore, the cluster would dissolve rapidly once the tide switches from being compressive, to the more likely extensive mode. This process therefore provides a unique clue to explain the discrepancy which exists between the expected Gaussian and the near-power-law mass function in the Antennae. The weak dependance of clusters half-light radius with mass \citep{vandenBergh08} implies that compressive tides should be operative on all mass scales and severely imprint the mass function. As an aside, we note that the general character of compressive tides means that they are also at work at high-redshift, when merger rates were higher.

\section{Conclusion}

Using gravitational N-body simulation of the Antennae galaxy merger, we show that a significant fraction of the discs' mass undergoes totally compressive tides. This mode  plays an important role by preventing a cluster from dissolving. The compressive regions detected in the simulations show a very good match with observational data, both in term of their spatial distribution, and their characteristic duration which compare to observed clusters ages.

Compressive modes of the tidal field help keep together clusters that would otherwise dissolve at formation, through e.g. rapid gas expulsion. By introducing a tidal term in the virial Eq. (\ref{eq:a}), we found that clusters should remain bound when they sit inside a compressive tide, even when the SFE is as low as 10\%. The precise characteristics of the evolution of clusters is clearly dependent on the relative strength of the tidal field. This is a hint that to understand the survival of clusters, and especially the issues associated with the concept of ``infant mortality'', it is essential to include the full evolution of the environment in which clusters evolve. These super-virial young clusters sheltered by compressive tides will dissolve rapidly whenever the tides are no longer compressive. Hydrodynamical simulations will be required to link formally this effect with cluster formation histories.

\section*{Acknowledgments}
We thanks Mark Gieles for thoughtful comments. FR acknowledges a scholarship from the IK~I033-N \emph{Cosmic Matter Circuit} at the University of Vienna. TN and CT are grateful for financial support within a common DFG project on the Antennae system granted by the DFG priority program 1177.

\end{document}